\documentclass[prd,twocolumn,showpacs,superscriptaddress,nofootinbib]{revtex4-1}
\usepackage[T1]{fontenc} % if needed
\usepackage{amsmath}
\usepackage{amssymb}
\usepackage{amsfonts}
\usepackage{graphicx}
\usepackage{enumitem}
\usepackage{bm}
\usepackage{rotating}
\usepackage{hyperref}
\usepackage{xcolor}
\usepackage{array}
\usepackage{lmodern}
\usepackage[normalem]{ulem}
\usepackage{braket}
\usepackage{tensor}
\usepackage{mathrsfs}
\usepackage{float}

\usepackage[normalem]{ulem}

\begin{document}

\title{$U(1)_{Y'}^{}$ universal seesaw }

\author{Su-Ping Chen}

\email{spchen@seu.edu.cn}

\author{Pei-Hong Gu}

\email{phgu@seu.edu.cn}

\affiliation{School of Physics, Jiulonghu Campus, Southeast University, Nanjing 211189, China}

\begin{abstract}

We extend the $SU(3)_c^{} \times SU(2)_L^{} \times U(1)_Y^{}$ standard model by a $U(1)_{Y'}^{}$ gauge symmetry. Three right-handed neutrinos are introduced to cancel the gauge anomaly. One Higgs singlet is responsible for spontaneously breaking the $U(1)_{Y'}^{}$ symmetry while the standard model Higgs doublet does not carry any $U(1)_{Y'}^{}$ charges. The down-type quarks, up-type quarks, charged leptons and neutral neutrinos obtain their Dirac masses through four types of dimension-5 operators constructed by the fermion doublets and singlets with the Higgs doublet and singlet. This effective theory is realized in three renormalizable contexts with heavy fermion singlets, scalar doublets and fermion doublets. The heavy fermion singlets and doublets for generating the neutrino masses also accommodate a successful Dirac leptogenesis to explain the baryon asymmetry in the universe.

\end{abstract}

%\pacs{14.80.Va, 14.80.Ec, 14.65.Jk, 12.60.Fr}

\maketitle

\section{Introduction} 

In the $SU(3)_c^{} \times SU(2)_L^{} \times U(1)_Y^{}$ standard model (SM) of particle physics, the charged fermions obtain their Dirac masses through the dimension-4 Yukawa interactions among the left-handed fermion doublets, the right-handed fermion singlets and the Higgs doublet. On the other hand, the right-handed neutrinos are absent from the SM. Consequently the neutrinos keep massless in the SM. However, the discovery of neutrino oscillations firmly indicate that the neutrinos should be massive and mixing \cite{workman2022}. This phenomenon calls for new physics beyond the SM. The simplest way for generating the neutrino masses seems to introduce the right-handed neutrinos and then construct the dimension-4 Yukawa interactions, like the scheme for generating the charged fermion masses. Unfortunately, the cosmological observations stringently constrain the neutrinos to be ultralight \cite{workman2022}. This fact means the Yukawa couplings of the neutrinos to the Higgs doublet should be extremely small, even if they are compared with that of the lightest charged fermion, i.e. the electron \cite{workman2022}.

In order to naturally explain the tiny but nonzero neutrino masses, ones proposed the elegant seesaw mechanism \cite{minkowski1977,yanagida1979,grs1979,ms1980,mw1980,sv1980,cl1980,lsw1981,ms1981,flhj1989} where the SM is extended by certain lepton number violating terms involving heavy particles, including the right-handed neutrino singlets with heavy Majorana masses, the Higgs triplets with small vacuum expectation values (VEVs), and the left-handed fermion triplets with heavy Majorana masses. In this seesaw context, the right-handed neutrinos, the Higgs triplets and the fermion triplets indeed mediate a Weinberg dimension-5 operator \cite{weinberg1979} constructed by the left-handed lepton doublets and the Higgs doublet. The lepton number violation and then the Majorana neutrinos are just a theoretical assumption and have not been verified in any experiments yet \cite{workman2022}. So, ones applied the seesaw mechanism to the Dirac neutrinos \cite{rw1983,rs1984,mp2002,gh2006,gu2012}, in analogy to the Majorana neutrino case. In these Majorana or Dirac seesaw models, the cosmic baryon asymmetry, which is another big challenge to the SM, can be understood in a natural way \cite{fy1986,dlrw1999}. This is the famous leptogenesis mechanism \cite{fy1986}.

Needless to say, the neutrino mass generation is so different from the charged fermion mass generation in the framework of the SM plus its seesaw extension. This somewhat mysterious feature may be addressed in a universal seesaw scenario \cite{berezhiani1983,rajpoot1987,dw1987} where not only the neutrinos but also the charged fermions acquire their masses through the seesaw mechanism. An early attempt was to consider the $SU(3)_c^{}\times SU(2)_L^{}\times SU(2)_R^{}\times U(1)_{B-L}^{}$ left-right symmetric model \cite{ps1974,mp1975,sm1975} with minimal Higgs sector including a left-handed doublet and its right-handed partner \cite{berezhiani1983,rajpoot1987,dw1987}. In this unconventional left-right symmetric scenario, the SM left-handed fermions are still placed in the $SU(2)_L^{}$ doublets, while the SM right-handed charged fermions and the right-handed neutrinos are placed in the $SU(2)_R^{}$ doublets. Then the left-handed and right-handed fermion and Higgs doublets can construct some dimension-5 operators mediated by additionally heavy fermion singlets with appropriate electric charges \cite{berezhiani1983,rajpoot1987,dw1987}.

In this work we extend the SM $SU(3)_c^{} \times SU(2)_L^{} \times U(1)_Y^{}$ gauge symmetry by a $U(1)_{Y'}^{}$ gauge symmetry \cite{bgm1991,langacker2008}. According to three generations of the SM fermions, we introduce three right-handed neutrinos to cancel the gauge anomaly. The Higgs sector only contains a Higgs singlet and a Higgs doublet. The Higgs singlet is responsible for spontaneously breaking the $U(1)_{Y'}^{}$ symmetry while the Higgs doublet does not carry any $U(1)_{Y'}^{}$ charges and is just the SM one. The down-type quarks, up-type quarks, charged leptons and neutral neutrinos obtain their Dirac masses through four types of dimension-5 operators constructed by the fermion doublets and singlets with the Higgs doublet and singlet. We then realize this effective theory in three renormalizable contexts with heavy fermion singlets, scalar doublets and fermion doublets. The heavy fermion singlets and doublets for generating the neutrino masses also accommodate a successful neutrinogenesis mechanism \cite{dlrw1999}, i.e. the Dirac leptogenesis to explain the baryon asymmetry in the universe.

\section{Effective theory}

\begin{table}
%\vspace{0.25cm}
\begin{center}
\begin{tabular}{|l|c|c|c|c|c|}  \hline &&&&\\[-2.0mm] ~~$Fermion~\&~Higgs$~~&~$SU(3)_c^{}$~&~$SU(2)_L^{}$~&~$U(1)_Y^{}$~&~$U(1)_{Y'}^{}$~\\
&&&&\\[-2.0mm]\hline&&&& \\[-1.5mm]
~~~~~~~~~$q_L^{}=\left[\begin{array}{c}u_L^{}\\
[2mm]
d_L^{}\end{array}\right]$&$3$ & $2$ & $+\frac{1}{6}$ &$-\frac{1}{4}$  \\
&&&&\\[-2.0mm]\hline&&&& \\[-1.5mm]~~~~~~~~~$d_R^{}$ &$3$ &$1$ & $-\frac{1}{3}$& $-\frac{3}{4}$  \\
&&&&\\[-2.0mm]\hline&&&& \\[-1.5mm]~~~~~~~~~$u_R^{}$ &$3$  & $1$& $+\frac{2}{3}$ &$+\frac{1}{4}$ \\
&&&&\\[-2.0mm]\hline&&&& \\[-1.5mm]~~~~~~~~~$l_L^{}=\left[\begin{array}{c}\nu_L^{}\\
[2mm]
e_L^{}\end{array}\right]$ &$1$  & $2$& $-\frac{1}{2}$ &$+\frac{3}{4}$\\
&&&&\\[-2.0mm]\hline&&&& \\[-1.5mm]~~~~~~~~~$e_R^{}$ &$1$  & $1$& $-1$ &$+\frac{1}{4}$ \\
&&&&\\[-2.0mm]\hline&&&& \\[-1.5mm]~~~~~~~~~$\nu_R^{}$ &$1$  & $1$& ~$0$ &$+\frac{5}{4}$ \\
&&&&\\[-2.0mm]\hline&&&& \\[-1.5mm]~~~~~~~~~$\phi=\left[\begin{array}{c}\phi^{+}_{}\\
[2mm]
\phi^{0}_{}\end{array}\right]$ &$1$  & $1$& $+\frac{1}{2}$ &~$0$ \\
&&&&\\[-2.0mm]\hline&&&& \\[-1.5mm]~~~~~~~~~$\xi$ &$1$  & $1$& ~$0$ &$+\frac{1}{2}$ \\
[1.5mm]
\hline
\end{tabular}
\vspace{0.25cm}
\caption{\label{fields} The related fermions and scalars for the universal seesaw within the context of effective theory. Here $q_L^{}$, $d_R^{}$, $u_R^{}$ respectively are the SM left-handed quark doublets, right-handed down-type quarks, right-handed up-type quarks, $l_L^{}$ and $e_R^{}$ respectively are the SM left-handed lepton doublets and right-handed charged leptons, while $\nu_R^{}$ are the non-SM right-handed neutrinos. The family indices of fermions have been omitted for simplicity. As the Higgs singlet $\xi$ is responsible for the $U(1)_{Y'}^{}$ symmetry breaking, the Higgs doublet $\phi$ does not carry any $U(1)_{Y'}^{}$ charges and drives the $SU(2)_L^{}\times U(1)_Y^{}$ symmetry breaking. }
\end{center}
\end{table}

\begin{figure*}
\centering
\includegraphics[scale=0.68]{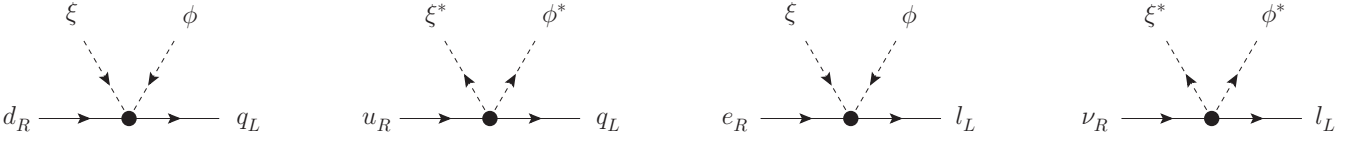}
\caption{The dimension-5 operators for generating the Dirac masses of up-type quarks, down-type quarks, charged leptons, and neutral neutrinos. }
\label{dim-5}
\end{figure*}

In this section we demonstrate the universal seesaw within the context of effective theory. The related fermions and scalars are summarized in Table {\ref{fields}. The Higgs doublet $\phi$ only carries the SM quantum numbers and is still responsible for the electroweak symmetry breaking. The new Higgs singlet $\xi$ drives the $U(1)_{Y'}^{}$ symmetry breaking. Besides the three generations of the SM fermions, i.e. $q_L^{}$, $d_R^{}$, $u_R^{}$, $l_L^{}$ and $e_R^{}$, the fermion sector contains three right-handed neutrinos $\nu_R^{}$. Here and thereafter we have omitted the family indices for simplicity.

Actually, for the SM fermions and the right-handed neutrinos, their $Y'$ quantum numbers can be well determined by $Y'=Y-\frac{5}{4}(B-L)$ \cite{bgm1991,langacker2008} with $Y$, $B$ and $L$ being the hypercharge, baryon number and lepton number, respectively. Since the SM Higgs doublet $\phi$ is not charged under the global symmetries of baryon and lepton number, it usually can carry a $Y'$ charge $Y'=+\frac{1}{2}$, as same as its hypercharge. In this case, the dimension-4 Yukawa couplings are available as usual, 
\begin{eqnarray}
\label{yukawa}
\mathcal{L}  &\supset & -y_d^{}\bar{q}_L^{} \phi d_R^{} - y_u^{} \bar{q}_L^{}\tilde{\phi} u_R^{} - y_e^{} \phi e_R^{} - y_\nu^{} \bar{l}_L^{} \tilde{\phi} \nu_R^{} + \textrm{H.c.}\,.\nonumber\\
&&
 \end{eqnarray}

However, the SM Higgs doublet definitely need not obey the relation $Y'=Y-\frac{5}{4}(B-L)$ since only the SM quarks and leptons have a specific definition of baryon and lepton number. So, we here do not assign any $Y'$ charges to the SM Higgs doublet as shown in Table \ref{fields}. In consequence, the dimension-4 Yukawa couplings (\ref{yukawa}) are exactly forbidden. Instead, the fermion doublets and singlets can construct the following dimension-5 operators with the SM Higgs doublet and the new Higgs singlet, i.e.
\begin{eqnarray}
\label{eff}
\mathcal{L}^{}&\supset& - \frac{c_d^{}}{\Lambda_d^{}}\bar{q}_L^{}\phi d_R^{} \xi - \frac{c_u^{}}{\Lambda_u^{}}\bar{q}_L^{}\tilde{\phi} u_R^{} \xi^\ast_{} - \frac{c_e^{}}{\Lambda_e^{}}\bar{e}_L^{}\phi e_R^{} \xi \nonumber\\
&&
-\frac{c_\nu^{}}{\Lambda_\nu^{}}\bar{l}_L^{} \tilde{\phi} \nu_R^{} \xi^\ast_{}+\textrm{H.c.}\,.
\end{eqnarray}
This effective theory can be also understood in Fig. \ref{dim-5}.

When the new and electroweak symmetries are broken spontaneously, the Higgs singlet $\xi$ and the Higgs doublet $\phi$ can be expressed by 
\begin{eqnarray}
\xi=\frac{1}{\sqrt{2}}\left(v_\xi^{}+h_\xi^{}\right)\,,~~\phi=\frac{1}{\sqrt{2}}\left[\begin{array}{c} 0\\
[2mm]
v_\phi^{}+h_\phi^{}\end{array}\right]\,,
 \end{eqnarray}
with $v_{\xi,\phi}^{}$ and $h_{\xi,\phi}^{}$ being the vacuum expectation values (VEVs) and Higgs bosons respectively. The fermions then can acquire their Dirac masses, i.e. 
\begin{eqnarray}
m_d^{} &=& \frac{c_d^{}v_\xi^{}v_\phi^{}}{2\Lambda_d^{}}\,,~~m_u^{}= \frac{c_u^{}v_\xi^{}v_\phi^{}}{2\Lambda_u^{}}\,,~~m_e^{}= \frac{c_e^{}v_\xi^{}v_\phi^{}}{2\Lambda_e^{}}\,,\nonumber\\
m_\nu^{}&= &\frac{c_\nu^{}v_\xi^{}v_\phi^{}}{2\Lambda_\nu^{}}\,.
 \end{eqnarray}
Once the new VEV $v_\xi^{}$ is fixed, the hierarchy among the above fermion masses can be elegantly understood by choosing either large cutoff $\Lambda_{d,u,e,\nu}^{}$ or small couplings $c_{d,u,e,\nu}^{}$.

Note due to our chosen charge assignments in Table \ref{fields}, the famous Weinberg dimension-5 operator for generating the Majorana neutrino masses should be absent from the present effective theory, i.e.
\begin{eqnarray}
\mathcal{L}  ~~&/\!\!\!\!\!\supset &~ -\frac{c'^{}_\nu}{2\Lambda'^{}_\nu}\bar{l}_L^{}\tilde{\phi}\tilde{\phi}^T_{}l_L^c+ \textrm{H.c.} ~~\Rightarrow~~m'^{}_\nu =\frac{c'^{}_\nu v_\phi^{2}}{2\Lambda_\nu^{}} \,.
 \end{eqnarray}

\section{Renormalizable models}

In this section we construct the renormalizable models with heavy fermion singlets, scalar doublets or fermion doublets to realize the effective theory (\ref{eff}).

\subsection{Heavy fermion singlet case}

\begin{table}
%\vspace{0.25cm}
\begin{center}
\begin{tabular}{|l|c|c|c|c|c|}  \hline &&&&\\[-2.0mm] ~~$Fermion~singlets$~~&~$SU(3)_c^{}$~&~$SU(2)_L^{}$~&~$U(1)_Y^{}$~&~$U(1)_{Y'}^{}$~\\
&&&&\\[-2.0mm]\hline&&&& \\[-1.5mm]~~~~~~~~~~$D_{L,R}^{}$&$3$ & $1$ & $-\frac{1}{3}$ &$-\frac{1}{4}$  \\
&&&&\\[-2.0mm]\hline&&&& \\[-1.5mm]~~~~~~~~~~$U_{L,R}^{}$ &$3$ &$1$ & $+\frac{2}{3}$& $-\frac{1}{4}$  \\
&&&&\\[-2.0mm]\hline&&&& \\[-1.5mm]~~~~~~~~~~$E_{L,R}^{}$ &$1$  & $1$& $-1$ &$+\frac{3}{4}$ \\
&&&&\\[-2.0mm]\hline&&&& \\[-1.5mm]~~~~~~~~~~$N_{L,R}^{}$ &$1$  & $1$& ~$0$ &$+\frac{3}{4}$ \\
[1.5mm]
\hline
\end{tabular}
\vspace{0.25cm}
\caption{\label{fsinglets}  The four types of heavy vector-like fermion singlets with $U(1)_{Y'}^{}$ charge. }
\end{center}
\end{table}

\begin{figure*}
\centering
\includegraphics[scale=0.68]{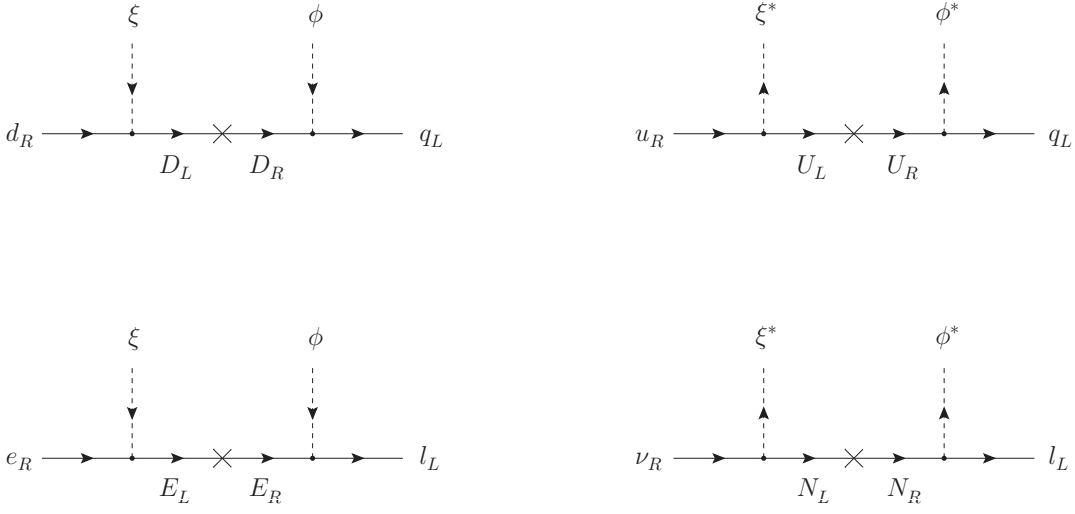}
\caption{The four types of heavy vector-like fermion singlets respectively mediate the dimension-5 operators for generating the Dirac masses of up-type quarks, down-type quarks, charged leptons, and neutral neutrinos. }
\label{singlet}
\end{figure*}

We now demonstrate the renormalizable models with heavy vector-like fermion singlets. See Table \ref{fsinglets}. The Yukawa and mass terms involving the vector-like fermion singlets should be 
\begin{eqnarray}
\label{lfsinglet}
\mathcal{L}^{}&\supset& - y_D^{} \bar{q}_L^{}\phi D_R^{} - f_D^{}\bar{D}_L^{} d_R^{}\xi - \hat{M}_D^{} \bar{D}_L^{} D_R^{}\nonumber\\
&&- y_U^{} \bar{q}_L^{}\tilde{\phi} U_R^{} - f_U^{}\bar{U}_L^{} u_R^{}\xi^\ast_{} - \hat{M}_U^{} \bar{U}_L^{} U_R^{}\nonumber\\
&& - y_E^{} \bar{l}_L^{}\phi E_R^{} - f_E^{}\bar{E}_L^{} e_R^{}\xi - \hat{M}_E^{} \bar{E}_L^{} E_R^{}\nonumber\\
&&- y_N^{} \bar{l}_L^{}\tilde{\phi} N_R^{} - f_N^{}\bar{N}_L^{} \nu_R^{}\xi^\ast_{} - \hat{M}_N^{} \bar{N}_L^{} N_R^{}+\textrm{H.c.}\,.
\end{eqnarray}
Here the mass matrices are chosen to be real and diagonal without loss of generality and for convenience. It should be noted that the electric neutral fermion singlets $N_{L,R}^{}$ are forbidden to have any Majorana masses because of their $U(1)_{Y'}^{}$ charge. This is different from the left-right symmetric models \cite{dw1987}.

If the vector-like fermion singlets are heavy enough, they can be integrated out, as shown in Fig. \ref{singlet}. The dimension-5 operators (\ref{eff}) then can be given by 
\begin{eqnarray}
\mathcal{L}^{}&\supset& \left(y_D^{}\frac{1}{\hat{M}_D^{}}f_D^{} \right)\bar{q}_L^{}\phi d_R^{}\xi + \left(y_U^{}\frac{1}{\hat{M}_U^{}}f_U^{} \right)\bar{q}_L^{}\tilde{\phi} u_R^{}\xi^\ast_{} \nonumber\\
&& + \left(y_E^{} \frac{1}{\hat{M}_E^{}} f_E^{}\right)\bar{l}_L^{}\phi e_R^{}\xi + \left( y_N^{} \frac{1}{\hat{M}_N^{}}f_N^{}\right)\bar{l}_L^{}\tilde{\phi}  \nu_R^{}\xi^\ast_{} \nonumber\\
&& +\textrm{H.c.}\,.
\end{eqnarray}
Accordingly, the Yukawa couplings in Eq. (\ref{yukawa}) are determined by 
\begin{eqnarray}
y_d^{}&=&- y_D^{}\frac{v_\xi^{}}{\sqrt{2}\hat{M}_D^{}}f_D^{} \,,~~y_u^{}=-y_U^{}\frac{v_\xi^{}}{\sqrt{2}\hat{M}_U^{}}f_U^{} \,,\nonumber\\
 y_e^{}&=&-y_E^{} \frac{v_\xi^{}}{\sqrt{2}\hat{M}_E^{}} f_E^{}\,,~~ y_\nu^{}=-y_N^{} \frac{v_\xi^{}}{\sqrt{2}\hat{M}_N^{}}f_N^{}\,,
\end{eqnarray}
which can be suppressed either by the products of two Yukawa couplings or by the ratio between VEV and mass \cite{berezhiani1983,rajpoot1987,dw1987,cm1987,babumoh1989,babumoh1989-2}. As a result, the Yukawa coupling parameters can have "more natural" values compared to their values in the SM. This is most noticeable for the Dirac neutrinos, which in the SM would have required $y_\nu^{}\sim 10^{-12}_{}$ whereas due to seesaw property, we need $y_N^{}\sim f_N^{}\sim 10^{-6}$ for $\hat{M}_N^{}\sim v_\xi^{}$.

Note when the above effective Yukawa couplings are the unique source of the SM fermion masses, they need at least three generations of heavy $D$, $U$ and $E$ to guarantee three generations of massive $d$, $u$ and $e$. As for three generations of $\nu$, they have at least two massive eigenstates and hence requires at least two generations of heavy $N$.

\subsection{Heavy scalar doublet case}

\begin{table}
%\vspace{0.25cm}
\begin{center}
\begin{tabular}{|l|c|c|c|c|c|}  \hline &&&&\\[-2.0mm] ~~$Higgs~doublets$~~&~$SU(3)_c^{}$~&~$SU(2)_L^{}$~&~$U(1)_Y^{}$~&~$U(1)_{Y'}^{}$~\\
&&&&\\[-2.0mm]\hline&&&& \\[-1.5mm]~~~~~$\eta=\left[\begin{array}{c}\eta^{+}_{}\\
[2mm]
\eta^{0}_{}\end{array}\right]$&$1$ & $2$ & $+\frac{1}{2}$ &$+\frac{1}{2}$   \\
[5mm]
\hline
\end{tabular}
\vspace{0.25cm}
\caption{\label{hdoublets}  The heavy scalar doublets with $U(1)_{Y'}^{}$ charge.}
\end{center}
\end{table}

\begin{figure*}
\centering
\includegraphics[scale=0.68]{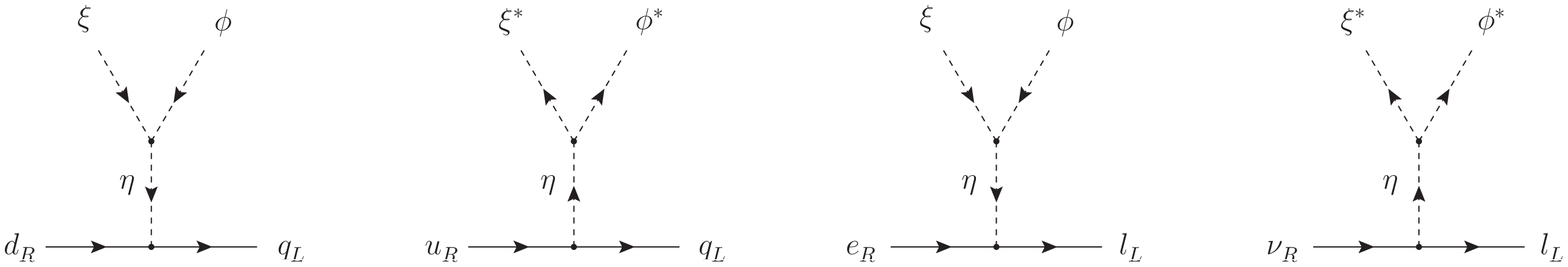}
\caption{The same heavy scalar doublets mediate the dimension-5 operators for generating the Dirac masses of up-type quarks, down-type quarks, charged leptons, and neutral neutrinos.  }
\label{higgsdoublet}
\end{figure*}

The heavy scalar doublets $\eta$ defined in Table \ref{hdoublets} can also mediate the dimension-5 operators (\ref{eff}), as shown in Fig. \ref{higgsdoublet}. The relevant Lagrangian includes
\begin{eqnarray}
\label{lhdoublet}
\mathcal{L}^{}&\supset&-\hat{M}_\eta^2 \eta^\dagger_{}\eta - \mu_\eta^{} \xi \eta^\dagger_{} \phi - f_d^{} \bar{q}_L^{} \eta d_R^{} - f_u^{} \bar{q}_L^{} \tilde{\eta} u_R^{} \nonumber\\
&&- f_e^{} \bar{l}_L^{} \eta e_R^{} - f_\nu^{} \bar{l}_L^{} \tilde{\eta}\nu_R^{} +\textrm{H.c.}\,.
\end{eqnarray}
For two or more heavy scalar doublets, their mass matrix $\hat{M}_{\eta}^2$ is taken to be real and diagonal without loss of generality and for convenience. For the same reason, their cubic couplings $\mu_\eta^{}$ are also taken to be real.

By integrating out the heavy scalar doublets $\eta$ from Eq. (\ref{lhdoublet}), we can obtain the dimension-5 operators (\ref{eff}), i.e. 
\begin{eqnarray}
\mathcal{L}^{}&\supset& \left(f_d^{}\frac{1}{\hat{M}_\eta^{2}}\mu_\eta^{}\right)\bar{q}_L^{}\phi d_R^{}\xi + \left(f_u^{}\frac{1}{\hat{M}_\eta^{2}} \mu_\eta^{}\right)\bar{q}_L^{}\tilde{\phi} u_R^{}\xi^\ast_{} \nonumber\\
&& + \left(f_e^{} \frac{1}{\hat{M}_\eta^{2}} \mu_\eta^{}\right)\bar{l}_L^{}\phi e_R^{}\xi + \left( f_\nu^{}\frac{1}{\hat{M}_\eta^{2}}\mu_\eta^{}\right)\bar{l}_L^{}\tilde{\phi}  \nu_R^{}\xi^\ast_{} \nonumber\\
&& +\textrm{H.c.}\,.
\end{eqnarray}
and then the Yukawa couplings (\ref{yukawa}), i.e. 
\begin{eqnarray}
y_d^{}&=&- f_d^{}\frac{v_\xi^{}}{\sqrt{2}\hat{M}_\eta^{2}}\mu_\eta^{} \,,~~y_u^{}=-f_u^{}\frac{v_\xi^{}}{\sqrt{2}\hat{M}_\eta^{2}}\mu_\eta^{} \,,\nonumber\\
 y_e^{}&=&-f_e^{} \frac{v_\xi^{}}{\sqrt{2}\hat{M}_\eta^{2}} \mu_\eta^{}\,,~~ y_\nu^{}=-f_\nu^{} \frac{v_\xi^{}}{\sqrt{2}\hat{M}_\eta^{2}}\mu_\eta^{}\,.
\end{eqnarray}
It is straightforward to see
\begin{eqnarray}
y_d^{}:y_u^{}:y_e^{}:y_\nu^{}=f_d^{} : f_u^{}:f_e^{}:f_\nu^{}\,.
\end{eqnarray}
This means we should take $f_\nu^{}\ll f_{d,u,e}^{}$ for $m_\nu^{}\ll m_{d,u,e}^{}$. This definitely is not a natural explanation for the large hierarchy between the charged fermions and the neutral neutrinos. So we will no longer consider the heavy scalar doublets in the following.

\subsection{Heavy fermion doublet case}

\begin{table}
%\vspace{0.25cm}
\begin{center}
\begin{tabular}{|l|c|c|c|c|c|}  \hline &&&&\\[-2.0mm] ~~~~$Fermion~doublets$~~&\,$SU(3)_c^{}$\,&\,$SU(2)_L^{}$\,&\,$U(1)_Y^{}$\,&\,$U(1)_{Y'}^{}$\,\\
&&&&\\[-2.0mm]\hline&&&& \\[-1.5mm]~$\Omega_{L}^{}=\!\left[\begin{array}{c}\Omega_L^{+\frac{2}{3}}\\
[2mm]
\Omega_L^{-\frac{1}{3}}\end{array}\right]$&$3$ & $2$ & $+\frac{1}{6}$ &$-\frac{3}{4}$  \\
&&&&\\[-2.0mm]\hline&&&& \\[-1.5mm]~$\Omega_R^{}=\!\left[\begin{array}{c}\Omega_R^{+\frac{2}{3}}\\
[2mm]
\Omega_R^{-\frac{1}{3}}\end{array}\right]\!=i\tau_2^{} \Omega'^{c}_{L}$ &$3$ &$2$ & $+\frac{1}{6}$& $-\frac{3}{4}$  \\
&&&&\\[-2.0mm]\hline&&&& \\[-1.5mm]~$\Psi_{L}^{}=\!\left[\begin{array}{c}\Psi_L^{+\frac{2}{3}}\\
[2mm]
\Psi_L^{-\frac{1}{3}}\end{array}\right]$ &$3$  & $2$& $+\frac{1}{6}$ &$+\frac{1}{4}$ \\
&&&&\\[-2.0mm]\hline&&&& \\[-1.5mm]~$\Psi_R^{}=\!\left[\begin{array}{c}\Psi_R^{+\frac{2}{3}}\\
[2mm]
\Psi_R^{-\frac{1}{3}}\end{array}\right]\!=i\tau_2^{} \Psi'^{c}_{L}$ &$3$  & $2$& $+\frac{1}{6}$ &$+\frac{1}{4}$ \\
&&&&\\[-2.0mm]\hline&&&& \\[-1.5mm]~$\Sigma_{L}^{}=\!\left[\begin{array}{c}\Sigma_L^{0}\\
[2mm]
\Sigma_L^{-}\end{array}\right]$&$1$ & $2$ & $-\frac{1}{2}$ &$+\frac{1}{4}$  \\
&&&&\\[-2.0mm]\hline&&&& \\[-1.5mm]~$\Sigma_R^{}=\!\left[\begin{array}{c}\Sigma_R^{0}\\
[2mm]
\Sigma_R^{-}\end{array}\right]\!=i\tau_2^{} \Sigma'^{c}_{L}$ &$1$ &$2$ & $-\frac{1}{2}$& $+\frac{1}{4}$  \\
&&&&\\[-2.0mm]\hline&&&& \\[-1.5mm]~$\Delta_{L}^{}=\!\left[\begin{array}{c}\Delta_L^{0}\\
[2mm]
\Delta_L^{-}\end{array}\right]$ &$1$  & $2$& $-\frac{1}{2}$ &$+\frac{5}{4}$ \\
&&&&\\[-2.0mm]\hline&&&& \\[-1.5mm]~$\Delta_R^{}=\!\left[\begin{array}{c}\Delta_R^{0}\\
[2mm]
\Delta_R^{-}\end{array}\right]\!=i\tau_2^{} \Delta'^{c}_{L}$ &$1$  & $2$& $-\frac{1}{2}$ &$+\frac{5}{4}$ \\
[5mm]
\hline
\end{tabular}
\vspace{0.25cm}
\caption{\label{fermiondoublets}  The four types of heavy vector-like fermion doublets with $U(1)_{Y'}^{}$ charge.}
\end{center}
\end{table}

\begin{figure*}
\centering
\includegraphics[scale=0.68]{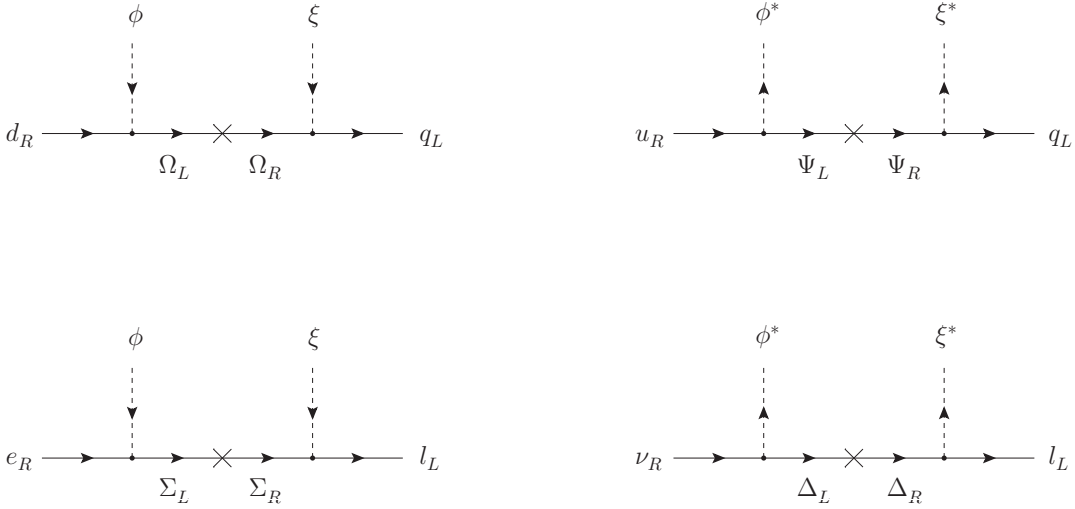}
\caption{The four types of heavy vector-like fermion doublets mediate the dimension-5 operators for generating the Dirac masses of up-type quarks, down-type quarks, charged leptons, and neutral neutrinos.  }
\label{doublet}
\end{figure*}

In analogy to the heavy vector-like fermion singlet case, we can realize the dimension-5 operators (\ref{eff}) through the mediation of heavy vector-like fermion doublets, as shown in Fig. \ref{doublet}. The heavy vector-like fermion doublets are defined in Table \ref{fermiondoublets}. Their Yukawa and mass terms are given by 
\begin{eqnarray}
\label{lfdoublet}
\mathcal{L}^{}&\supset& - y_\Omega^{} \bar{q}_L^{}\Omega_R^{}\xi - f_\Omega^{}\bar{\Omega}_L^{} \phi d_R^{} - \hat{M}_\Omega^{} \bar{\Omega}_L^{} \Omega_R^{}\nonumber\\
&&- y_\Psi^{} \bar{q}_L^{}\Psi_R^{}\xi^\ast_{} - f_\Psi^{}\bar{\Psi}_L^{} \tilde{\phi} u_R^{} - \hat{M}_\Omega^{} \bar{\Psi}_L^{} \Psi_R^{}\nonumber\\
&& - y_\Sigma^{} \bar{l}_L^{}\Sigma_R^{}\xi - f_\Sigma^{}\bar{\Sigma}_L^{} \phi e_R^{} - \hat{M}_\Sigma^{} \bar{\Sigma}_L^{} \Sigma_R^{}\nonumber\\
&&- y_\Delta^{} \bar{l}_L^{} \Delta_R^{}\xi^\ast_{} - f_\Delta^{}\bar{\Delta}_L^{} \tilde{\phi} \nu_R^{}  - \hat{M}_\Delta^{} \bar{\Delta}_L^{} \Delta_R^{}+\textrm{H.c.}\,.
\end{eqnarray}
Here the mass matrices are chosen to be real and diagonal without loss of generality and for convenience.

We integrate out the heavy vector-like fermion doublets $\Omega$, $\Psi$, $\Sigma$ and $\Delta$ from Eq. (\ref{lfdoublet}) to derive 
\begin{eqnarray}
\mathcal{L}^{}&\supset& \left(y_\Omega^{}\frac{1}{\hat{M}_\Omega^{}}f_\Omega^{} \right)\bar{q}_L^{}\phi d_R^{}\xi + \left(y_\Psi^{}\frac{1}{\hat{M}_\Psi^{}}f_\Psi^{} \right)\bar{q}_L^{}\tilde{\phi} u_R^{}\xi^\ast_{} \nonumber\\
&& + \left(y_\Sigma^{} \frac{1}{\hat{M}_\Sigma^{}} f_\Sigma^{}\right)\bar{l}_L^{}\phi e_R^{}\xi + \left( y_\Delta^{} \frac{1}{\hat{M}_\Delta^{}}f_\Delta^{}\right)\bar{l}_L^{}\tilde{\phi}  \nu_R^{}\xi^\ast_{} \nonumber\\
&& +\textrm{H.c.}\,.
\end{eqnarray}
and then 
\begin{eqnarray}
y_d^{}&=&- y_\Omega^{}\frac{v_\xi^{}}{\sqrt{2}\hat{M}_\Omega^{}}f_\Omega^{} \,,~~y_u^{}=-y_\Psi^{}\frac{v_\xi^{}}{\sqrt{2}\hat{M}_\Psi^{}}f_\Psi^{} \,,\nonumber\\
 y_e^{}&=&-y_\Sigma^{} \frac{v_\xi^{}}{\sqrt{2}\hat{M}_\Sigma^{}} f_\Sigma^{}\,,~~ y_\nu^{}=-y_\Delta^{} \frac{v_\xi^{}}{\sqrt{2}\hat{M}_\Delta^{}}f_\Delta^{}\,.
\end{eqnarray}
Note when the above effective Yukawa couplings are the unique source of the SM fermion masses, there should be at least three generations of heavy $\Omega$, $\Psi$ and $\Sigma$ for three generations of $d$, $u$ and $e$, meanwhile, at least two generations of heavy $\Delta$ for at least two nonzero mass eigenvalues of $\nu$.

\subsection{Top quark mass}

We should also keep in mind that the universal seesaw probably is not suitable for the top quark. This is because the seesaw condition would result in a too small mass to the top quark, i.e.  $\hat{M}_{U}^{} \gg y_{U}^{} v_{\phi}^{}/\sqrt{2}\,, f_{U}^{} v_{\xi}^{}/\sqrt{2}$ in the heavy fermion singlet case and $\hat{M}_{\Psi}^{} \gg f_{U}^{} v_{\phi}^{}/\sqrt{2}\,, y_{U}^{} v_{\xi}^{}/\sqrt{2}$ in the heavy fermion doublet case. However, we could consider a weaker seesaw condition for the top quark, i.e. $\hat{M}_{U}^{}\sim  f_{U}^{} v_{\xi}^{}/\sqrt{2}  \gg y_{U}^{} v_{\phi}^{}/\sqrt{2}$ and $\hat{M}_{\Psi}^{} \sim  y_{U}^{} v_{\xi}^{}/\sqrt{2} \gg f_{U}^{} v_{\phi}^{}/\sqrt{2}$. In this case, while the up component $u_{L3}^{}$ of the left-handed doublet $q_{L3}^{}$ rather than the left-handed singlets $U_{L3}^{}$ dominates the SM left-handed top quark $t^{}_L$, the two right-handed singlets $u_{R3}^{}$ and $U_{R3}^{}$ have a sizable mixing and then one of their two linear combinations become the SM right-handed top quark $t^{}_R$. Such arrangement should be acceptable as long as the fermion singlet $U_{3}^{}$ is heavy enough to suppress its correction to the CKM matrix. Similarly, the up components $\Psi_{L3}^{+\frac{2}{3}}$ and $u_{L3}^{}$ of the left-handed doublets $\Psi_{L3}^{}$ and $q_{L3}^{}$ have a sizable mixing to form the SM left-handed top quark $t_L^{}$.

\section{Baryon asymmetry}

In this section we demonstrate the heavy vector-like fermion singlets and doublets for suppressing the Dirac neutrino masses can account for the cosmological baryon asymmetry through their CP-violation and out-of-equilibrium decays. Specifically these decays can produce a lepton asymmetry stored in the left-handed lepton doublets and an opposite lepton asymmetry stored in the right-handed neutrinos. The lepton asymmetry in the left-handed lepton doublets can be partially converted to a baryon asymmetry by the $SU(2)_L^{}$ sphaleron processes \cite{krs1985} while the lepton asymmetry in the right-handed neutrinos keeps alone. This is because the effective Yukawa interactions between the left-handed lepton doublets and the right-handed neutrinos are extremely weak so that they can not go into equilibrium before the sphalerons stop working, unlike the other SM charged fermions \cite{dlrw1999}. 

\subsection{Fermion singlet decays}

\begin{figure*}
\centering
\includegraphics[scale=0.68]{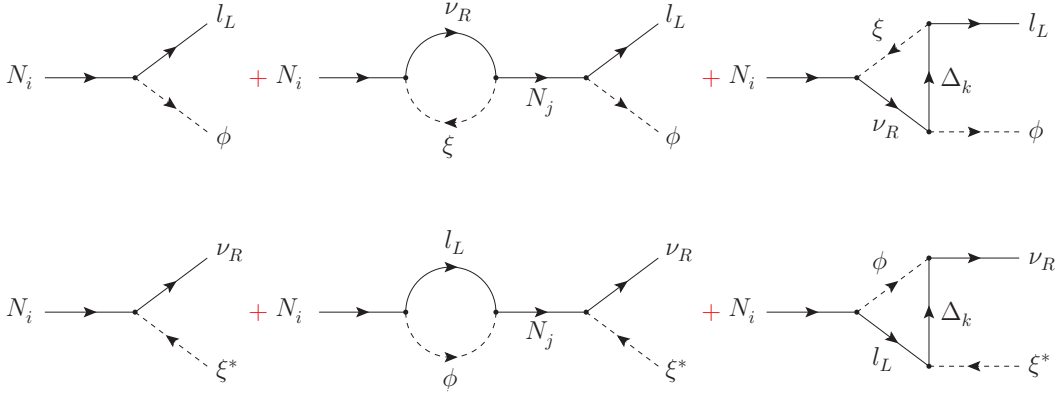}
\caption{The heavy vector-like fermion singlet decays at one-loop order. }
\label{FSdecay}
\end{figure*}

As shown in Fig. \ref{FSdecay}, the heavy vector-like fermion singlets have two decay modes,
\begin{eqnarray}
N_i^{}\rightarrow  l_L^{} + \phi\,,~~N_i^{} \rightarrow \nu_R^{}+ \xi^\ast_{}\,.
\end{eqnarray}
As long as the CP is not conserved, we can expect a CP asymmetry in the above decays, 
\begin{eqnarray}
\varepsilon_{N_i}^{} &=&\frac{\Gamma(N_i^{}\rightarrow  l_L^{} + \phi)- \Gamma(N_i^{c}\rightarrow  l_L^{c} + \phi^\ast_{})}{\Gamma_{N_i}^{}}\nonumber\\
& \equiv& \frac{ \Gamma(N_i^{c} \rightarrow \nu_R^{c}+ \xi) - \Gamma(N_i^{} \rightarrow \nu_R^{}+ \xi^\ast_{})}{\Gamma_{N_i}^{}} \neq 0\,,
\end{eqnarray}
where the total decay width $\Gamma_{N_i}^{}$ is given by 
\begin{eqnarray}
\Gamma_{N_i}^{}&=&\Gamma(N_i^{}\rightarrow  l_L^{} + \phi)+ \Gamma(N_i^{} \rightarrow \nu_R^{}+ \xi^\ast_{})\nonumber\\
&=&\Gamma(N_i^{c}\rightarrow  l_L^{c} + \phi^\ast_{})+ \Gamma(N_i^{c} \rightarrow \nu_R^{c}+ \xi)\,.
\end{eqnarray}

We calculate the decay width at tree level, i.e. 
\begin{eqnarray}
\label{widthn}
\Gamma_{N_i}^{}=\frac{1}{32\pi}\left[2\left(y_N^\dagger y_N^{}\right)_{ii}^{}+\left(f_N^{}f_N^\dagger\right)_{ii}^{}\right]M_{N_i}^{}\,,
\end{eqnarray}
and the CP asymmetry at one-loop order, i.e.
\begin{eqnarray}
\label{cpan}
\varepsilon_{N_i}^{}&=&\frac{\textrm{Im}\left\{\sum_{j\neq i}^{}\left[\left(y_N^\dagger y_N^{} \right)_{ij}^{} \left(f_N^{}f_N^\dagger\right)_{ji}\right]\right\}}{4\pi\left[2\left(y_N^\dagger y_N^{}\right)_{ii}^{}+\left(f_N^{}f_N^\dagger\right)_{ii}^{}\right]}\frac{M_{N_i}^{}M_{N_j}^{}}{M_{N_i}^2 - M_{N_j}^2}\nonumber\\
&&+ \frac{\textrm{Im}\left\{\sum_{k}^{}\left[\left(y_N^\dagger y_\Delta^{} \right)_{ik}^{} \left(f_\Delta^{}f_N^\dagger\right)_{ki}\right]\right\}}{2\pi \left[2\left(y_N^\dagger y_N^{}\right)_{ii}^{}+\left(f_N^{}f_N^\dagger\right)_{ii}^{}\right]}\frac{M_{\Delta_k}^{}}{M_{N_i}^{}}\nonumber\\
&&\times \left[1-\left(1+\frac{M_{\Delta_k}^2}{M_{N_i}^2}\right)\ln \left(1+\frac{M_{N_i}^2}{M_{\Delta_k}^2}\right)\right]\,.
\end{eqnarray}
Here the first term in the CP asymmetry is from the self-energy correction while the second term is from the vertex correction, like those in the decays of the right-handed neutrinos with heavy Majorana masses \cite{fy1986,fps1995,fpsw1996,crv1996,pilaftsis1997}. For a nonzero self-energy correction, we need at least two heavy fermion singlets.

In the hierarchical case with
\begin{eqnarray}
M_{N_i}^2 \ll M_{N_j}^2, M_{\Delta_k}^2\,,
\end{eqnarray}
the CP asymmetry (\ref{cpan}) can be simplified to be 
\begin{eqnarray}
\label{cpan2}
\varepsilon_{N_i}^{}&\simeq&\frac{1}{2\pi}\frac{M_{N_i}^{}\textrm{Im}\left[\left(y_N^\dagger m_\nu^{} f_N^\dagger\right)_{ii}\right]  }{v_\xi^{}v_\phi^{} \left[2\left(y_N^\dagger y_N^{}\right)_{ii}^{}+\left(f_N^{}f_N^\dagger\right)_{ii}^{}\right]}\,.
\end{eqnarray}
Here and thereafter $m_\nu^{}$ denotes the Dirac neutrino mass matrix,
 \begin{eqnarray}
m_\nu^{}= -y_N^{} \frac{v_\xi^{}v_\phi^{}}{2\hat{M}_N^{}}f_N^{}-y_\Delta^{} \frac{v_\xi^{}v_\phi^{}}{2\hat{M}_\Delta^{}}f_\Delta^{}\,.
\end{eqnarray}
The simplified CP asymmetry (\ref{cpan2}) roughly can have an upper bound \cite{di2002,bdp2003},
\begin{eqnarray}
\label{cpan3}
\varepsilon_{N_i}^{}&\leq & \frac{1}{2\pi}\frac{M_{N_i}^{}\textrm{Im}\left[\left(y_N^\dagger m_\nu^{} f_N^\dagger\right)_{ii}\right]  }{2 v_\xi^{}v_\phi^{} \sqrt{2\left(y_N^\dagger y_N^{}\right)_{ii}^{}\left(f_N^{}f_N^\dagger\right)_{ii}^{}}}\nonumber\\
&\lesssim & \frac{1}{4\sqrt{2}\pi}\frac{M_{N_i}^{}m_{\nu}^{\textrm{max}} }{v_\xi^{}v_\phi^{} }\,.
\end{eqnarray}
with $m_{\nu}^{\textrm{max}}$ being the maximal eigenvalue of the neutrino mass matrix $m_\nu^{}$.

\subsection{Fermion doublet decays}

\begin{figure*}
\centering
\includegraphics[scale=0.68]{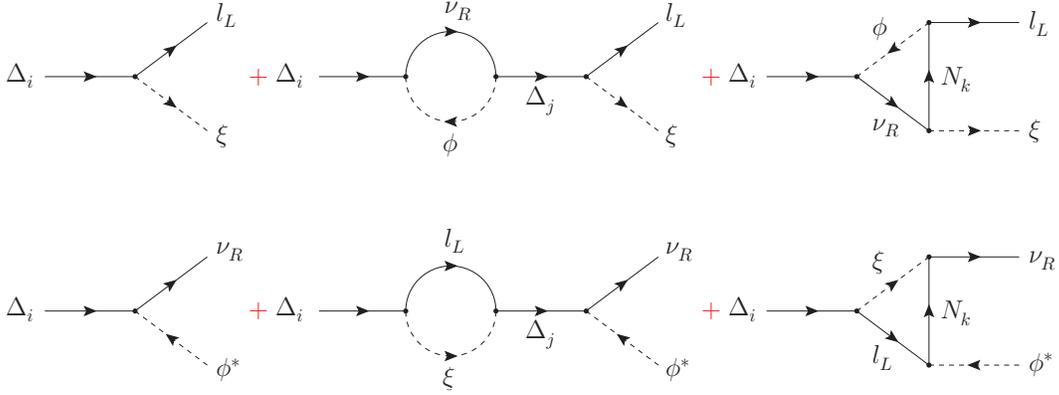}
\caption{ The heavy vector-like fermion doublet decays at one-loop order.}
\label{FDdecay}
\end{figure*}

The heavy vector-like fermion doublets also have two decay modes,
\begin{eqnarray}
\Delta_i^{}\rightarrow  l_L^{} + \xi\,,~~\Delta_i^{} \rightarrow \nu_R^{}+ \phi^\ast_{}\,.
\end{eqnarray}
See Fig. \ref{FDdecay}, where we can obtain a CP asymmetry as long as the CP is not conserved, i.e.
\begin{eqnarray}
\varepsilon_{\Delta_i}^{} &=&\frac{\Gamma(\Delta_i^{}\rightarrow  l_L^{} + \xi)- \Gamma(\Delta_i^{c}\rightarrow  l_L^{c} + \xi^\ast_{})}{\Gamma_{\Delta_i}^{}}\nonumber\\
& \equiv& \frac{ \Gamma(\Delta_i^{c} \rightarrow \nu_R^{c}+ \phi) - \Gamma(\Delta_i^{} \rightarrow \nu_R^{}+ \phi^\ast_{})}{\Gamma_{\Delta_i}^{}} \neq 0\,.
\end{eqnarray}
Here $\Gamma_{\Delta_i}^{}$ is the total decay,  
\begin{eqnarray}
\Gamma_{\Delta_i}^{}&=&\Gamma(\Delta_i^{}\rightarrow  l_L^{} + \xi)+ \Gamma(\Delta_i^{} \rightarrow \nu_R^{}+ \phi^\ast_{})\nonumber\\
&=&\Gamma(\Delta_i^{c}\rightarrow  l_L^{c} + \xi^\ast_{})+ \Gamma(\Delta_i^{c} \rightarrow \nu_R^{c}+ \phi)\,.
\end{eqnarray}

While the decay width is given at tree level, 
\begin{eqnarray}
\label{widthd}
\Gamma_{\Delta_i}^{}=\frac{1}{32\pi}\left[\left(y_\Delta^\dagger y_\Delta^{}\right)_{ii}^{}+\left(f_\Delta^{}f_\Delta^\dagger\right)_{ii}^{}\right]M_{\Delta_i}^{}\,.
\end{eqnarray}
the CP asymmetry should be calculated at one-loop order,
\begin{eqnarray}
\label{cpad}
\varepsilon_{\Delta_i}^{}&=&\frac{\textrm{Im}\left\{\sum_{j\neq i}^{}\left[\left(y_\Delta^\dagger y_\Delta^{} \right)_{ij}^{} \left(f_\Delta^{}f_\Delta^\dagger\right)_{ji}\right]\right\}}{8\pi\left[\left(y_\Delta^\dagger y_\Delta^{}\right)_{ii}^{}+\left(f_\Delta^{}f_\Delta^\dagger\right)_{ii}^{}\right]}\frac{M_{\Delta_i}^{} M_{\Delta_j}^{}}{M_{\Delta_i}^2 - M_{\Delta_j}^2 }\nonumber\\
&&+ \frac{\textrm{Im}\left\{\sum_{k}^{}\left[\left(y_\Delta^\dagger y_N^{} \right)_{ik}^{} \left(f_N^{}f_\Delta^\dagger\right)_{ki}\right]\right\}}{4\pi\left[\left(y_\Delta^\dagger y_\Delta^{}\right)_{ii}^{}+\left(f_\Delta^{}f_\Delta^\dagger\right)_{ii}^{}\right]}\frac{M_{N_k}^{}}{M_{\Delta_i}^{}}\nonumber\\
&&\times 
\left[1-\left(1+\frac{M_{N_k}^2}{M_{\Delta_i}^2}\right)\ln \left(1+\frac{M_{\Delta_i}^2}{M_{N_k}^2}\right)\right]\,.
\end{eqnarray}
Here the first term in the CP asymmetry is from the self-energy correction while the second term is from the vertex correction. For a nonzero self-energy correction, we need at least two heavy fermion doublets.

We then consider the hierarchical case with
\begin{eqnarray}
M_{\Delta_i}^2 \ll M_{\Delta_j}^2, M_{N_k}^2\,,
\end{eqnarray}
to simplify the CP asymmetry (\ref{cpad}) by
\begin{eqnarray}
\label{cpad2}
\varepsilon_{\Delta_i}^{}&\simeq &\frac{1}{4\pi}\frac{M_{\Delta_i}^{}\textrm{Im}\left[\left(y_\Delta^\dagger m_\nu^{} f_\Delta^\dagger\right)_{ii}\right]}{v_\xi^{}v_\phi^{}\left[\left(y_\Delta^\dagger y_\Delta^{}\right)_{ii}^{}+\left(f_\Delta^{}f_\Delta^\dagger\right)_{ii}^{}\right]}\nonumber\\
&\leq& \frac{1}{8\pi}\frac{M_{\Delta_i}^{}\textrm{Im}\left[\left(y_\Delta^\dagger m_\nu^{} f_\Delta^\dagger\right)_{ii}\right]}{v_\xi^{}v_\phi^{}\sqrt{\left(y_\Delta^\dagger y_\Delta^{}\right)_{ii}^{}\left(f_\Delta^{}f_\Delta^\dagger\right)_{ii}^{}}}\nonumber\\
&\lesssim&\frac{1}{8\pi}\frac{M_{\Delta_i}^{}m_\nu^{\textrm{max}} }{v_\xi^{}v_\phi^{}}\,.
 \end{eqnarray}

\subsection{Final baryon asymmetry}

After the heavy vector-like fermion singlets $N_i^{}$ or the heavy vector-like fermion doublets $\Delta_i^{}$ go out of equilibrium, their decays can generate a lepton number $L_{l_L}^{}=L$ stored in the left-handed lepton doublets $l_L^{}$ and an opposite lepton number $L_{\nu_R}^{}=-L$ stored in the right-handed neutrinos $\nu_R^{}$. For simplicity, we can assume a hierarchical spectrum of the heavy decaying particles, i.e. $M_{N_1}^{2}\ll M_{N_{2,...}}^{2},M_{\Delta_{2,...}}^2$, or $M_{\Delta_1}^{2}\ll M_{\Delta_{2,...}}^{2},M_{N_{1,...}}^2$. In this case, the decays of the lightest $N_1^{}/\Delta_1^{}$ should dominate the final $L$, i.e.
\begin{eqnarray}
L&=& \varepsilon_{N_1/\Delta_1}^{}\left(\frac{n^{eq}_{N_1/\Delta_1} }{s}\right)\left|_{T=T_D^{}}^{}\right.\,.
\end{eqnarray}
Here the symbols $n^{eq}_{N_1/\Delta_1}$ and $T_D^{}$ respectively are the equilibrium number density and the decoupled temperature of the heavy decaying fermions $N_1^{}/\Delta_1^{}$, while the character $s$ is the entropy density of the universe \cite{kt1990}. The sphaleron processes eventually will partially transfer this lepton number to a baryon asymmetry \cite{ht1990},
\begin{eqnarray}
\label{bauf}
B= -\frac{28}{79}L=-\frac{28}{79} \varepsilon_{N_1/\Delta_1}^{}\left(\frac{n^{eq}_{N_1/\Delta_1} }{s}\right)\left|_{T=T_D^{}}^{}\right.\,.
\end{eqnarray}
In principle the decoupled temperature $T_D^{}$ and then the final baryon number $B$ can be numerically solved by the related Boltzmann equations \cite{kt1990} which are beyond the scope of the present work.

\section{Effective neutrino number}

The right-handed neutrinos will affect the effective neutrino number which is stringently constrained by the BBN \cite{workman2022}. We hope the right-handed neutrinos can decouple above the QCD scale and hence give a negligible contribution to the effective neutrino number \cite{kt1990}. For this purpose, we need check the annihilations of the right-handed neutrinos into the other light species,  
\begin{eqnarray}
\sigma_{\nu_R}^{} &=&\sum_{f=d,u,s,e,\mu,\nu_L^{}}^{}\sigma(\nu_R^{} +\bar{\nu}_R^{}\rightarrow f+\bar{f}) \nonumber\\
&=& \frac{2825}{3\times 2^{12}_{}\pi}\frac{g_{Y'}^4}{M_{Z'}^4}s =  \frac{2825}{3\times 2^8_{}\pi}\frac{s}{v_\xi^4}\,.
\end{eqnarray}
Here $s$ is the Mandelstam variable, $g_{Y'}^{}$ is the $U(1)_{Y'}^{}$ gauge coupling, $Z'$ is the $U(1)_{Y'}^{}$ gauge boson and has a mass as below, 
\begin{eqnarray}
M_{Z'}^2=\frac{1}{4}g_{Y'}^2 v_\xi^2\,.
\end{eqnarray}
The interaction rate then should be \cite{gnrrs2003}
\begin{eqnarray}
\Gamma_{\nu_R}^{} &=&\frac{\frac{T}{32\pi^4_{}}\int^{\infty}_{0} s^{3/2}_{} K_1^{}\left(\frac{\sqrt{s}}{T}\right) \sigma_{\nu_R}^{}ds }{\frac{2}{\pi^2_{}}T^3_{}}= \frac{2825}{64\pi^3_{}} \frac{T^5_{}}{v_\xi^4}\,,
\end{eqnarray}
with $K_1^{}$ being a Bessel function.

Comparing the above interaction rate with the Hubble constant, 
\begin{eqnarray}
H(T)=\left[\frac{8\pi^{3}_{}g_{\ast}^{}(T)}{90}\right]^{\frac{1}{2}}_{}\frac{T^2_{}}{M_{\textrm{Pl}}^{}}\,,
\end{eqnarray}
we can find 
\begin{eqnarray}
\label{bbn}
\Gamma_{\nu_R^{}}^{} < H(T)\left|_{T\simeq 300\,\textrm{MeV}}^{}\right.~~\Longrightarrow~~v_\xi^{} \gtrsim 14\,\textrm{TeV}\,.
\end{eqnarray}
Here $M_{\textrm{Pl}}^{}\simeq 1.22\times 10^{19}_{}\,\textrm{GeV}$ is the Planck mass and $g_{\ast}^{}(T)\simeq 61.75$ is the relativistic degrees of freedom at $T\simeq 300\,\textrm{MeV}$ \cite{kt1990}. The bound (\ref{bbn}) could be safe for the other experimental constraints \cite{workman2022,langacker2008,klq2016,ccdm2019,dbhm2021}.

\section{Summary}

In this paper we have demonstrated the universal seesaw scenario based on the $SU(3)_c^{} \times SU(2)_L^{} \times U(1)_Y^{} \times U(1)_{Y'}^{}$ gauge groups. With this $U(1)_{Y'}^{}$ symmetry, we can naturally introduce the right-handed neutrinos to cancel the gauge anomaly. In the popular $U(1)_{Y'}^{}$ models \cite{bgm1991}, the SM Higgs doublet carries the $U(1)_{Y'}^{}$ charge and hence the usual dimension-4 Yukawa couplings are still allowed, meanwhile, a non-SM Higgs singlet is responsible for spontaneously breaking the $U(1)_{Y'}^{}$ symmetry and for generating the Majorana masses of the right-handed neutrinos. In our scenario, the SM Higgs doublet does not carry any $U(1)_{Y'}^{}$ charges so that the SM fermions and the right-handed neutrinos can not have the usual Yukawa couplings. Instead, the down-type quarks, up-type quarks, charged leptons and neutral neutrinos can obtain their Dirac masses through four types of dimension-5 operators constructed by the fermion doublets and singlets with the Higgs doublet and singlet. We then explore the renormalizable context with heavy fermion singlets, Higgs doublets and fermion doublets to realize these effective operators. The heavy fermion singlets or doublets for neutrino mass generation can accommodate a successful Dirac leptogenesis to explain the baryon asymmetry in the universe.

\textbf{Acknowledgement}: This work was supported in part by the National Natural Science Foundation of China under Grant No. 12175038 and in part by the Fundamental Research Funds for the Central Universities.

\end{document}